\documentstyle[12pt]{article}
\newcommand{\beq}{\begin{equation}}
\newcommand{\eeq}{\end{equation}}
\newcommand{\ben}{\begin{eqnarray}}
\newcommand{\een}{\end{eqnarray}}
\begin{document}
\begin{center}
{\Huge \bf Non-locally Regularized Field-Antifield Quantization of the
Chiral Schwinger Model} \\
\vspace{1cm}

{\large Everton M. C. Abreu\footnote{\noindent e-mail: everton@feg.unesp.br}}\\

%\end{center}
\vspace{1cm}

Instituto de F\'\i sica, Universidade Federal  do Rio de Janeiro,\\
Caixa Postal 68528, 21945  Rio de Janeiro,
RJ, Brazil\\

\end{center} 

\vspace{0.5cm}
\abstract

The non-local regularization is a powerfull method to regularize theories
with an action that can be decomposed in
a kinetic and an interacting part.
Recently it was shown how to regularize the Batalin-Vilkovisky 
field-antifield formalism 
of quantization of gauge theories with the non-local regularization.
We compute precisely the anomaly of the Chiral Schwinger model with
this extended non-local regularization.


\vfill\eject

\section{Introduction}

The method developed by Batalin-Vilkovisky (BV)\cite{BV} showed itself to
be a very powerfull way to quantize the most difficult field theories.  
A two dimensional gauge theory, the string theory, is one of these examples.  
For a review see \cite{Jon,Gomis,Hen}.

The BV, or field-antifield, formalism provides, at lagrangian level, a general 
framework for the covariant path integral quantization of gauge theories.  
This formalism uses very interesting mathematical objects like
a Poisson-like bracket (the antibracket), canonical 
transformations, ghosts for the BRST transformations, etc.  The most important 
object of this method at the classical level is an equation called 
classical master equation (CME).

The fundamental idea of the formalism is BRST invariance.  The fields 
$\Phi^{A}$, i.e., the classical fields of the theory toghether with the 
ghosts, and the auxiliary fields, 
have their canonically conjugated, the antifields $\Phi^{*}_{A}$.  
With all this 
elements we construct the so called BV action.  At the classical level, 
the BV action becomes the classical action when all the antifields are put 
to be zero.  A gauge-fixed action can be obtained by a canonical 
transformation.  At this time we can say that the action is in a gauge-fixed 
basis.  The other way to fix the gauge is towards a choice of a gauge 
fermion  
and making the antifields to be equal to the functional derivative of this 
fermion.  The method can be applied to gauge theories which have an open 
algebra (the algebra of gauge transformations closes only on shell),
to closed algebras, to gauge 
theories that have structure functions rather than constants (soft algebras), 
and
to the case where the gauge transformations may or may not be independent, 
reducible or irreducible algebras respectively.

Zinn-Justin introduced the concept of sources of BRST-transformations 
\cite{ZJ}.  These sources are the antifields in the BV formalism.  It was 
shown also that the geometry of the antifields have a natural origin 
\cite{Wit}.

At quantum level, the field-antifield formalism also works at one-loop 
anomalies \cite{Troost,Pro}.  There, with the addition of extra degrees of 
freedon, causing an extension of the original configuration space, we
have 
a solution for a regularized quantum master equation (QME) at one-loop,
that has 
been obtained as a part independent of the antifields in the anomaly.  When 
the Wess-Zumino terms, which cancell the anomaly,
can not be found, the theory can be said to have a 
genuine anomaly.  Recently, a method was developed to handle with global 
anomalies \cite{Nelson}, when a quantity that is conserved classically is 
not conserved at quantum level.

However, the solution of the QME is not easily obtained because there is a 
divergence when the $\Delta$ operator, a two order differential operator 
defined below, is applied on local functionals, 
a $\delta(0)$-like divergence.  Therefore, a 
regularization method has to be used to cut the divergence
 in the QME.  One of them is the Pauli-Villars (PV) regularization 
\cite{Pauli,Diaz,Hat}, where new fields, the PV fields, and an arbitrary mass 
matrix are introduced.  But this method is very usefull only at one-loop 
level.  At higher orders, the PV method is still misterious.  Very recently, a 
BPHZ renormalization \cite{BPHZ} of the BV formalism was formulated 
\cite{Jonghe,EU}.  A dimensional regularization method in the quantum aspect 
of the field-antifield quantization has been studied in ref. \cite{Tonin}.

The non-local regularization (NLR) \cite{NL,Kle,Woo} gives a consistent way to 
compute anomalies at higher order levels of $\hbar$.  The main ideas were 
based in Schwinger's proper time method \cite{Sch}.  The 
preliminary results \cite{Hand,Clay} were very well received. The  
NLR separates the 
original divergent loop integrals in a sum over loop contribution in such a 
way that the loops, now composed of a set of auxiliary fields, contain the 
original singularities.  To regularize the original theory one has to 
eliminate these auxiliary fields by putting them on shell.  In this way the 
theory is free of the quantum fluctuations.  An extension of the NLR
method to the BV framework has been recently formulated by J. Paris 
\cite{Paris}.

In this work we regularize the chiral Schwinger model (CSM), which has been
constructed and completely solved by Jackiw and Rajaraman \cite{JR}, 
within the 
context of this extended 
non-local BV regularization.  The anomaly at one-loop is computed precisely.  
In section 2 a brief review of the field-antifield formalism has been made.
In section 3 the original NRL is depicted.  The extended non-local 
regularization is described in section 4.  The computation of the CSM anomaly 
at one-loop has been calculated in section 5.  The conclusions 
and final remarks were accomplished in section 6.

\section{The Field-Antifield Formalism}

Let's construct the complete set of fields, including in this set 
the classical fields, the
ghosts for all gauge symmetries and the auxiliary fields.  This
complete set will be denoted by ${\Phi^{A}}$.  Now, let's extend this
space with the same number of fields, but at this time, one will
define the antifields ${\Phi_{A}^{*}}$, which is the canonical
conjugated variables with respect to the antibracket structure.  This
is constructed like
\beq
(X,Y) = \frac{\delta_{r} X}{\delta \phi}\,\frac{\delta_{l} Y}{\delta
\phi^{*}} - ( X \longleftrightarrow Y ),
\eeq
where the indices $r$ and $l$ denote right and left derivation
respectively. 

By means of antibrackets, one can write the canonical conjugation
relations 
\beq
\label{conj}
(\Phi^{A},\Phi^{*}_{A}) = \delta^{A}_{B}\,\,,\,\,\,
(\Phi^{A},\Phi^{B}) = (\Phi^{*}_{A},\Phi^{*}_{B}) = 0.
\eeq

The antifields $\Phi^{*}_{A}$ have opposite statistics than their
conjugated fields $\Phi^{A}$.  The antibracket is a fermionic operation
so that the statistics of the antibracket $(X,Y)$ is opposite to that
of $XY$.  The antibracket also satifies some graded Jacobi relations:
\beq
(X,(Y,Z)) +(-)^{\epsilon_{X}\epsilon_{Y} + \epsilon_{X} +
\epsilon_{Y}} (Y,(X,Z)) = ((X,Y),Z).
\eeq
where $\epsilon_{X}$ is the statistics of $X$, i.e. 
$\epsilon(X) = \epsilon_{X}$.

We define a quantity named ghost number to fields and antifields.
These are integers such that
\beq
gh(\Phi^{*}) = - 1 - gh(\Phi).
\eeq

One can then construct an action of ghost number zero so that it is an
extended action, the so called BV action, also called classical
proper solution, so that
\beq
\label{proaction}
S(\Phi,\Phi^{*}) = S_{0}(\Phi) + \Phi^{*}_{A} R^{A}(\Phi) +
\frac{1}{2}\Phi^{*}_{A} \Phi^{*}_{B} R^{AB}(\Phi) + \ldots +
\frac{1}{n!} \Phi^{*}_{A_{1}} \ldots \Phi^{*}_{A_{n}} R^{A_{n}\ldots
A_{1}} + \ldots
\eeq

This equation contains all the algebra of the theory, the gauge
invariances of the classical action $(S_{cl} =
S_{BV}(\Phi^{A},\Phi^{*}_{A}=0))$, Jacobi identities,...
Gauge fixing is obtained either by a canonical transformation or by
choosing a fermion $\Psi^{A}$ and writing 
\beq
\Phi^{*} = \frac{\delta_{r} \Psi^{A}}{\delta \Phi^{A}}
\eeq

At quantum level the quantum action can be defined by:
\beq
\label{exp}
W = S + \sum^{\infty}_{p=1} \hbar^{p} M_{p},
\eeq
where the $M_{i}$ are the quantum corrections, the Wess-Zumino terms,
to the quantum action.  The expansion (\ref{exp}) is not the only one, 
but is
the usual one.  An expansion in $\sqrt{\hbar}$ \cite{Sie}  can be made.
This will originate the so called background charges, that is usefull
in conformal field theory \cite{Back}.

The quantization of the theory is made by the Green function's
generating functional:
\beq
Z(J,\Phi^{*}) = \int {\cal D} \Phi \,exp \, \frac{i}{\hbar} \left[ W (\Phi,
\Phi^{*}) + J_{A} \Phi^{*A} \right] .
\eeq
But the definition of a path integral properly lacks on a
regularization framework, which can be seen as a way to define the
measure. Anomalies represent the non conservation of the classical
symmetries at quantum level.  
%Physically when the theory has an
%anomaly, the antifields become propagating fields.
%These are called induced theories.

For a theory to be free of anomalies, the quantum action $W$ has to
be a solution of the QME,
\beq
\label{qme}
(W,W) = 2\,i\,\hbar\, \Delta\, W
\eeq
where
\beq
\Delta \equiv (-1)^{A+1} \frac{\partial_{r}}{\partial \Phi^{A}}
\frac{\partial_{r}}{\partial\Phi^{*}_{A}}.
\eeq

In the equation (\ref{qme})  one can see that:
\beq
{\cal A} \equiv \left[ \Delta W + \frac{i}{2\hbar} (W,W) \right] (\Phi, \Phi^{*}).
\eeq
And computing a $\hbar$ expansion,
\beq
{\cal A} = \sum^{\infty}_{p=0} \hbar^{p-1} M_{p}
\eeq
one have the form of the $p$-loop BRST anomalies:
\ben
{\cal A}_{0} & = & \frac{1}{2}\,(S,S) \equiv 0 \\
{\cal A}_{1} & = & \Delta S + i\,(M_{1},S) \\
{\cal A}_{2} & = & \Delta M_{p-1} +
\frac{i}{2}\sum^{p-1}_{q=1} (M_{q},M_{p-q}) + i(M_{p},S)\,\,, p \geq 2
\een

The first equation is the known CME.  The second one is an equation
for $M_{1}$.  If the second equation does not have a solution for
$M_{1}$ then ${\cal A}$ is called anomaly.  The anomaly is not
uniquely determined since $M_{1}$ is arbitrary.  The anomaly satisfy
the Wess-Zumino consistency condition \cite{WZ}:
\beq
({\cal A},S) = 0.
\eeq

%If we split the equation (\ref{qme}) in powers of $\hbar$, the two first
%equations, the zero-loop and the one-loop are:
%\ben
%(S,S) & = & 0 \\
%i {\cal A} & \equiv & i \Delta S - (M_{1},S) = 0.
%\een

\section{The Non-local Regularization}

As we explained at the introduction, the non-local regularization can
be applied only to theories which have a perturbative expansion,
i.e. for actions that can be decomposed into a free and an
interacting part.  For much more details, including the diagramatic
part, the interested reader can see the references \cite{NL,Kle,Woo,Paris}
\footnote{\noindent For convenience we are using the same notation of the
reference \cite{Paris}.}.

Let's define an action $S(\Phi)$ where $\Phi$ is the set $\Phi^{A}$
of the fields, $A=1,\ldots,N$, and with statistics
$\epsilon(\Phi^{A}) \equiv \epsilon_{A}$.
\beq
\label{origaction}
S(\Phi) = F(\Phi) + I(\Phi),
\eeq
$F(\Phi)$ is the kinetic part and $I(\Phi)$ is the interacting part.

Then one can write
\beq
F(\Phi) = \frac{1}{2} \Phi^{A} {\cal F}_{AB} \Phi^{B}
\eeq
and $I(\Phi)$ is an analitic function in $\Phi^{A}$ around
$\Phi^{A} = 0$. ${\cal F}_{AB}$ is called the kinetic
operator. 

We have now to introduce a cut-off or regulating parameter
$\Lambda^{2}$.  An arbitrary and invertible matrix $T_{AB}$ has to
be introduced too.  With the combination between ${\cal F}_{AB}$ and 
$(T^{-1})^{AB}$ we can define a second order derivative regulator:
\beq
{\cal R}^{A}_{B} = (T^{-1})^{AC} {\cal F}_{AB}.
\eeq

The construction of two important operators can be made with these
objects.  The first is the smearing operator
\beq
\epsilon^{A}_{B} = exp \left( \frac{{\cal R}^{A}_{B}}{2\Lambda^{2}} \right) ,
\eeq
and the second is the shadow kinetic operator
\beq
{\cal O}_{AB}^{-1} = T_{AC}(\tilde{{\cal O}}^{-1})^{C}_{B} = 
\left( \frac{{\cal F}}{\epsilon^{2} - 1} \right)_{AB},
\eeq
with $(\tilde{{\cal O}})^{A}_{B}$ defined as
\beq
\tilde{\cal O}^{A}_{B} = \left( \frac{\epsilon^{2} - 1}{{\cal R}} \right)^{A}_{B}
= \int_{0}^{1} \frac{dt}{\Lambda^{2}}\, 
exp \left(\,t\, \frac{{\cal R}^{A}_{B}}{\Lambda^{2}} \right).
\eeq

Expanding our original configuration space, for each field
$\Phi^{A}$ an auxiliary field $\Psi^{A}$ has been constructed, the
shadow field, with the same statistics.  A new auxiliary action
couple both sets of fields
\beq
\label{auxaction}
\tilde{{\cal S}}(\Phi,\Psi) = F(\hat{\Phi}) - A(\Psi) + I(\Phi + \Psi).
\eeq
The second term of this auxiliary action is called kinetic term, and
is constructed by
\beq
A(\Psi) = \frac{1}{2}\Psi^{A} ({\cal O}^{-1})_{AB} \Psi^{B}.
\eeq
The fields $\hat{\Phi}^{A}$, the smeared fields, which make part of
the auxiliary action are defined by
\beq
\hat{\Phi}^{A} \equiv (\epsilon^{-1})^{A}_{B} \Phi^{B}.
\eeq

It can be proved that, to eliminate the quantum fluctuations
associated with the shadow fields at the path integral level one has to
accomplish this by puting the auxiliary fields $\Psi$ on shell.  So, the
classical shadow fields equations of motion are
\beq
\label{claeq}
\frac{\partial_{r} \tilde{S}(\Phi,\Psi)}{\partial \Psi} = 0
\Longrightarrow \Psi^{A} =\left( \frac{\partial_{r} I}{\partial
\Phi^{B}}(\Phi + \Psi) \right) {\cal O}^{BA}.
\eeq
These equations can be solved in a perturbative fashion.  The
classical solutions $\bar{\Psi}_{0}(\Phi)$ can now be substituted in
the auxiliary action (\ref{auxaction}).  This substitution modify the
auxiliary action so that a new action, the non-localized action appear,
\beq
\label{nlocaction}
{\cal S}_{\Lambda}(\Phi) \equiv \tilde{\cal S}(\Phi,\bar{\Psi}_{0}(\Phi)).
\eeq

The action (\ref{nlocaction}) can be expanded in $\bar{\Psi}_{0}$.
As a result, we see the appearance of the smeared kinetic term
$F(\hat{\Phi})$, the original interaction term $I(\Phi)$ and an
infinite series of new non-local interaction terms.  But all these
interaction terms are of $O\left(\frac{1}{\Lambda^{2}}\right)$ and
when the limit $\Lambda^{2} \longrightarrow \infty$ will be made, then we
will have that ${\cal S}_{\Lambda}(\Phi) \longrightarrow {\cal
S}(\Phi)$, and the original theory is obtained.  Equivalently to this
limit, the same result can be acquired with the limits
\beq
\epsilon \longrightarrow 1,\,\,\,\,\,\,\,\,\,\, {\cal O} \longrightarrow 0,
\,\,\,\,\,\,\,\,\,\, \bar{\Psi}_{0}(\Phi) \longrightarrow 0.
\eeq

With all this framework, when we introduce the smearing operator, any
local quantum field theory can be made ultraviolet finite.  But a
question about symmetry can appear.   Obviously this form of
non-localization destroy any kind of gauge symmetry or its associated
BRST symmetry.  The final consequence is the damage of the
corresponding Ward identities at the tree level.

Let's make an analysis of what happen.  If the original
action (\ref{origaction}) is invariant under the infinitesimal
transformation 
\beq
%\label{brst}
\delta\,\Phi^{A} = R^{A}(\Phi)
\eeq
so, the auxiliary action is invariant under the auxiliary
infinitesimal transformations
\ben
\label{brst}
\tilde{\delta} \Phi^{A} & = & \left( \epsilon^{2} \right)^{A}_{B}\,R^{B}\,
(\Phi + \Psi), \nonumber \\
\tilde{\delta} \Psi^{A} & = & \left( 1-\epsilon^{2} \right)^{A}_{B}\,
R^{B}\,(\Phi + \Psi).
\een

However, the non-locally regulated action (\ref{nlocaction}) is
invariant under the transformation
\beq
\delta_{\Lambda} (\Phi^{A}) = \left( \epsilon^{2} \right)^{A}_{B}\,R^{B} \left( \Phi + 
\bar{\Psi}_{0}(\Phi) \right),
\eeq
remembering that $\bar{\Psi}_{0}(\Phi)$ are the solutions of the
classical equations of motions (\ref{claeq}).

Hence, any of the original continuosus symmetries of the theory are
preserved at the tree level, even the BRST transformations, and
consequently, the
original gauge symmetry.  The reader can see \cite{NL,Kle,Woo} for
details. 

\section{The Extended (BV) Non-local Regularization}

As had been said before, the fundamental principle of the
field-antifield formalism is BRST invariance.  Therefore, it is 
simple to realize that the conection between the NLR method and the BV
formalism is possible.  Using the above construction of the NLR and the BV
results, one can build a regulated BRST classical structure of a
general gauge theory from the original one.  Consequently, a
non-locally regularized BV formalism comes out.

We are now in the BV environment.  Hence, the configuration space has
to be enlarged introducing the antifields $\{\Psi^{A},\Psi_{A}^{*}\}$.
Note that the shadow fields have antifields too.  Then, an auxiliary
proper solution, which incorporates the auxiliary
action (\ref{auxaction}), corresponding to the gauge-fixed action
$S(\Phi)$, its BRST symmetry (\ref{brst}) and the unknown
associated higher order structure functions.  The auxiliary BRST
transformations (\ref{brst}), are modified by the presence of the
term $\Phi^{*}_{A}\,R^{A}(\Phi)$ in the original proper solution.
Then it can be written that the BRST transformations are 
\beq
\left[ \Phi^{*}_{A}(\epsilon^{2})^{A}_{B} +
\Psi^{*}_{A}(1-\epsilon^{2})^{A}_{B} \right] \,R^{B}\,\left(\Phi +
\Psi \right)
\eeq
which are originated from the substitution
\ben
\label{subst}
\Phi^{*}_{A} & \longrightarrow & \left[ \Phi^{*}_{A}(\epsilon^{2})^{A}_{B} +
\Psi^{*}_{A}(1-\epsilon^{2})^{A}_{B} \right] \equiv \Theta^{*}_{A}
\nonumber \\
R^{A} & \longrightarrow &R^{A}(\Phi + \Psi) \equiv R^{A}(\Theta).
\een

For higher orders, the natural way would be
\beq
R^{A_{n}\ldots A_{1}}(\Phi) \longrightarrow
R^{A_{n}\ldots A_{1}}(\Phi+\Psi) = R^{A_{n}\ldots A_{1}}(\Theta)
\eeq
and an obvious ansatz for the auxiliary proper solution is
\ben
\label{ansataction}
\tilde{S}(\Phi,\Phi^{*};\Psi,\Psi^{*}) \, & = & \,\tilde{S}(\Phi,\Psi) \,+\,
\Theta^{*}_{A}\,R^{A}(\Theta) \, + \,
\Theta^{*}_{A}\Theta^{*}_{B}\,R^{AB}(\Theta) \, +  \nonumber \\
& + & \Theta^{*}_{A_{1}}\ldots \Theta^{*}_{A_{n}}\,R^{A_{n}\ldots A_{1}}(\Phi)
+ \ldots
\een

It is intuitive to see that the same canonical conjugation relations, 
equations (\ref{conj}), should be obtained, i.e.
\beq
\left( \Theta^{A},\Theta^{*}_{B} \right) = \delta^{A}_{B}.
\eeq
Consequently, we have to construct a new set of fields and antifields
$\{\Sigma^{A},\Sigma^{*}_{A}\}$ defined by
\beq
\Sigma^{A} = \left[ \left( 1-\epsilon^{2} \right)^{A}_{B}\Phi^{B} -
\left( \epsilon^{2} \right)^{A}_{B}\Psi^{B} \right],
\eeq
and
\beq
\Sigma^{*}_{A} = \Phi^{*}_{A} - \Psi^{*}_{A}.
\eeq

Now we have that the linear transformation
\beq
\{\Phi^{A},\Phi^{*}_{A};\Psi^{A},\Psi^{*}_{A}\} \longrightarrow 
\{\Theta^{A},\Theta^{*}_{A};\Sigma,\Sigma^{*}_{A}\}
\eeq
is canonical in the antibracket sense.  And the auxiliary
action (\ref{auxaction}) is the original proper
solution (\ref{proaction}) with arguments
$\{\Theta^{A},\Theta^{*}_{A}\}$.

The elimination of the auxiliaries fields in the BV method is the next
step.  The shadow fields have to be substituted by the solutions of
their classical equations of motion.  At the same time, their
antifields goes to zero.  In this way we can write
\beq
S_{\Lambda}(\Phi,\Phi^{*}) =
\tilde{S}(\Phi,\Phi^{*};\bar{\Psi},\Psi^{*} = 0),
\eeq
and the classical equations of motion are
\beq
\frac{\delta_{r}\,\tilde{S}(\Phi,\Phi^{*};\Psi,\Psi^{*})}
{\delta\,\Psi^{A}} = 0
\eeq
with solutions $\bar{\Psi} \equiv \bar{\Psi}(\Phi,\Phi^{*})$, which
explicitly read
\beq
\label{solution}
\bar{\Psi}^{A} = \left[ \frac{\delta_{r}\,I}{\delta \Phi^{B}}\,
\left(\Phi + \Psi \right)
+ \Phi^{*}_{C} \left( \epsilon^{2} \right)^{C}_{D} R^{D}_{B} 
\left( \Phi+\Psi \right) +
O \left( (\Phi^{*})^{2} \right) \right]
\eeq
with
\beq
R^{A}_{B} = \frac{\delta_{r}\,R^{A}\,(\Phi)}{\delta \Phi^{B}}.
\eeq
The lowest order of equation (\ref{solution}) is,
\beq
\bar{\Psi}^{A} = \left( \frac{\delta_{r}\,I}{\delta \Phi^{B}}(\Phi + \Psi)
\right) {\cal O}^{BA}
\eeq
and one can obtain an expression for $\bar{\Psi}(\Phi,\Phi^{*})$ at
any desired order in antifields \cite{Paris}.

To quantize the theory, it is necessary to add extra counterterms
$M_{p}$ to preserve the quantum counterpart of the classical BRST
scheme.  It is the same as to substitute the classical action $S$ by
a quantum action $W$.  In the original papers \cite{NL,Kle,Woo} 
the quantization of the theory
was already analyzed, but it seems that only one-loop $M_{1}$ 
corrections acquired
BRST invariance.  It can be proved that in the field-antifield
framework, in general, two and higher order loop corrections should
also be considered \cite{Paris}.

The complete interaction ${\cal I}(\Phi,\Phi^{*})$ of the original
proper solution can be written as
\beq
{\cal I}(\Phi,\Phi^{*}) \equiv I(\Phi) + \Phi^{*}_{A}\,R^{A}(\Phi) +
\Phi^{*}_{A}\,\Phi^{*}_{B}\,R^{AB}(\Phi) + \dots
\eeq
The non-localization of this interaction part furnishes a way to
regularize interactions from counterterms $M_{p}$.  To construct the
auxiliary free and interactions parts
\beq
\tilde{F}\,(\Phi + \Psi) = F(\hat{\Phi}) - A(\Psi),\,\,\,\,\,\,{\cal
I}\,(\Phi,\Phi^{*};\Psi,\Psi^{*}) = {\cal I}\,(\Theta,\Theta^{*})
\eeq
with $\{\Theta,\Theta^{*}\}$ already known.

Now one have to put the auxiliary fields on shell and its
antifields to zero, so that
\ben
F_{\Lambda}\,(\Phi,\Phi^{*}) & = & \tilde{F}\,(\Phi,\bar{\Psi}_{0}),
\nonumber \\
{\cal I}_{\Lambda}(\Phi,\Phi^{*}) & = & \tilde{{\cal
I}}\,(\Phi+\bar{\Psi}_{0},\Phi^{*} \epsilon^{2}),
\een
then $S_{\Lambda}=F_{\Lambda} + {\cal I}_{\Lambda}$.

The quantum action $W$ can be expressed by
\beq
W = F + {\cal I} + \sum_{p=1}^{\infty}\,\hbar\,M_{p} \equiv F + {\cal
Y}
\eeq
where ${\cal Y}$ now is the generalized quantum interaction.

An analogous procedure of the previous section can be applied to
the quantum action $W$.  We will omit all the formal steps here.

A decomposition in its divergent part and its finite part when
$\Lambda^{2} \longrightarrow \infty$ can be accomplished in the
regulated QME.

It can be shown that the expression of the anomaly is the value of
the finite part in the limit $\Lambda^{2} \longrightarrow \infty$ of 
\beq
\label{anomalia}
{\cal A} = \left[ (\,\Delta\,W\,)_{R} + \frac{i}{2\,\hbar}\,(W,W)
\right]\,(\Phi,\Phi^{*})
\eeq
and the regularized value of $\Delta W$ defined as
\beq
\label{operator}
(\Delta W)_{R} \equiv \lim_{\Lambda^{2} \rightarrow \infty} \left[
\Omega_{0} \right]
\eeq
where
\beq
\Omega_{0} = \left[
S_{B}^{A}\,\left( \delta_{\Lambda} \right)^{B}_{C}\, \left( \epsilon^{2} \right)_{A}^{C} \right].
\eeq

$\left( \delta_{\Lambda} \right)^{A}_{B}$ is defined by
\ben
(\delta_{\Lambda})_{B}^{A} & = & \left( \delta^{A}_{B} - {\cal
O}^{AC}\,{\cal I}_{CB} \right)^{-1} \nonumber \\
& = & \delta^{A}_{B} + \sum_{n=1}\, \left( {\cal O}^{AC}\,{\cal
I}_{CB} \right)^{n},
\een
with
\ben
S^{A}_{B} & = &
\frac{\delta_{r}\,\delta_{l}\,S}{\delta\,\Phi^{B}\,\delta\,\Phi^{*}_{A}}, \nonumber \\
{\cal I}_{AB} & = &
\frac{\delta_{r}\,\delta_{l}\,{\cal I}}{\delta\,\Phi^{A}\,\delta\,\Phi^{B}} 
\een

Applying the limit $\Lambda^{2} \longrightarrow \infty$ in 
(\ref{operator}), it can be shown that
\beq
\left( \Delta S\right)_{R} \equiv \lim_{\Lambda^{2} \rightarrow \infty}
\left[ \Omega_{0} \right]_{0}
\eeq

And finally that
\ben
{\cal A}_{0} & \equiv & \left( \Delta\,S \right)_{R} \nonumber \\
& = & \lim_{\Lambda^{2} \rightarrow \infty} \left[ \Omega_{0}
\right]_{0} 
\een

All the higher orders loop terms of the anomaly can be obtained from
equation (\ref{anomalia}), but this will not be analyzed in this paper.

\section{The Chiral Schwinger Model Extended Non-locally  Regularized}

The classical action for the chiral Schwinger model is
\beq
\label{csmaction}
S = \int\,d^{2}x \left[  - \frac{1}{4} \, F_{\mu\nu} F^{\mu\nu} +
\bar{\psi} i \not\partial \psi + {e \over 2}\bar{\psi}\gamma_{\mu} (1 -
\gamma_{5}) A^{\mu} \psi \right],
\eeq
which obviously has a perturbative expansion.

This action is invariant for the following  gauge transformations:
\ben
A^{\mu}(x) & \longrightarrow & A^{\mu}(x) + \partial_{\mu} \theta (x) \\
\psi(x) & \longrightarrow & exp\, \left[
i\,e\,(1-\gamma_{5})\,\theta(x)\, \right]\,\psi(x)
\een

The kinetic part of the action (\ref{csmaction}) is given by
\ben
F & = & \int d^{2}x\,\bar{\psi}\,i\not\partial \psi  \nonumber \\
& = & \int d^{2}x\, \left[ \frac{1}{2} \bar{\psi} \,i\not\partial \psi + 
\frac{1}{2} \bar{\psi} \,i \not\partial \psi \right]
\een

Integrating by parts the second term we have that
\beq
F = \int d^{2}x\, \left[ \frac{1}{2}\bar{\psi}\,i\not\partial \psi - 
\frac{1}{2}(i\not\partial^{t}\bar{\psi})\,\psi \right]
\eeq

The kinetic term has the form
\beq
F = \frac{1}{2} \Psi^{A}{\cal F}_{AB}\Psi^{B}
\eeq

So,
\beq
\Psi = \left( \begin{array}{c}
\bar{\psi} \\ \psi
\end{array} \right)
\eeq
and
\beq
F = \frac{1}{2}(\bar{\psi} \,\,\psi)\, 
\left( \begin{array}{cc}
0 & i\not\partial \\
i\not\partial^{t} & 0 
\end{array} \right) 
\left( \begin{array}{c}
\bar{\psi} \\ \psi 
\end{array} \right)
\eeq
and we have that the kinetic operator $( {\cal F}_{AB} )$ is
\beq
{\cal F}_{AB} = \left( \begin{array}{cc}
0 & i\not\partial \\
i\not\partial^{t} & 0 
\end{array} \right) 
\eeq

The regulator, a second order differential operator, is
\beq
{\cal R}^{\alpha}_{\beta} = (T^{-1})^{\alpha \gamma}{\cal F}_{\gamma \beta},
\eeq
where $T$ is an arbitrary matrix,
and one can make the following choice:
\beq
{\cal R}^{\alpha}_{\beta} = -\,\partial^{2}
\eeq

Using the definition of the smearing operator,
\beq
\epsilon^{A}_{B} = exp \left( \frac{- \partial^{2}}{2 \Lambda^{2}} \right),
\eeq
and the smeared fields are defined by
\beq
\hat{\Phi}^{A} = (\epsilon^{-1})^{A}_{B}\,\Phi^{B}
\eeq

In the NLR scheme the shadow kinetic operator is
\beq
{\cal O}_{\alpha\beta}^{-1} = 
\left( \frac{{\cal F}}{\epsilon^{2} - 1} \right)_{\alpha\beta}
\eeq
then
\beq
{\cal O} = \left( \begin{array}{cc}
0 & -i{\cal O}'\not\partial \\
-i{\cal O}'\not\partial^{t} & 0 
\end{array} \right) 
\eeq
where
\ben
{\cal O}' & = & \frac{\epsilon^{2} - 1}{\not\partial\not\partial^{t}} \nonumber \\
& = & \int_{0}^{1} \frac{dt}{\Lambda^{2}} exp
\left( t\, \frac{\not\partial^{t} \not\partial}{\Lambda^{2}} \right)
\een

The interacting part of the action (\ref{csmaction}) is
\ben
I\,\left[ A_{\mu},\psi,\bar{\psi} \right] & = &
e\,\bar{\psi}\,\gamma_{\mu}(1-\gamma_{5})A^{\mu}\,\psi \\
I \left[ A_{\mu},\psi+\Phi,\bar{\psi}+\bar{\Phi} \right] & = &
e\,( \bar{\psi} + \bar{\Phi} )\, \gamma_{\mu}(1-\gamma_{5})A^{\mu}\,
( \psi + \Phi )
\een
where $\Phi$ are the shadow fields.

The BRST transformations are given by
\ben
\delta A_{\mu} & = & \partial_{\mu}c, \nonumber \\
\delta \psi & = & i ( 1\,-\,\gamma_5 )\psi c, \nonumber \\
\delta\bar{\psi} & = & -\,i\bar{\psi} ( 1\,+\,\gamma_5 )c, \nonumber \\
\delta c & = & 0
\een

Making the substitution (\ref{subst}) where the antifields are 
functions of the auxiliary fields,
\ben
\psi^{*} \longrightarrow \left[ \psi^{*} \epsilon^{2} + \Phi^{*} (1-
\epsilon^{2}) \right] \nonumber \\
\bar{\psi}^{*} \longrightarrow \left[ \bar{\psi}^{*} \epsilon^{2} + 
\bar{\Phi}^{*} (1 - \epsilon^{2}) \right].
\een

The generator of BRST transformations are
\ben
R(\psi) & \longrightarrow & R(\psi + \Phi) = i ( 1\,-\,\gamma_5 )( \psi + \Phi) c \nonumber \\
R(\bar{\psi}) & \longrightarrow & i (\bar{\psi} + \bar{\Phi})( 1\,+\,\gamma_5 )c \nonumber \\
R(c) & = & 0
\een

We are able now to construct the non-local auxiliary proper action.
It will be given in general by
\beq
S_{\Lambda}(\Phi,\Phi^{*}) =
\tilde{S}_{\Lambda}(\Phi,\Phi^{*};\psi_{s},\psi^{*} = 0)
\eeq
where $\psi_{s}$ are the solutions of the classical equations of motion.

After an algebraic manipulation, one can write the non-localized
action as
\ben
\tilde{S}_{\Lambda}(\psi,\psi^{*}) & = & \hat{F}_{\mu\nu} \hat{F}^{\mu\nu} +
\hat{\bar{\psi}} i\not\partial \hat{\psi} + A^{*}_{\mu}\partial^{\mu}c +
\nonumber \\
& + & \frac{e\,(i\not\partial) \left[ \bar{\psi}\gamma^{\mu} (1 -
\gamma_{5}) A_{\mu} \psi \right]}{i\not\partial+e\,\gamma^{\mu} 
(1 -\gamma_{5}) A_{\mu}(\epsilon^{2}-1)}  + \nonumber \\
& + & \frac{i\,\psi^{*}\epsilon^{2}c(-i\not\partial) \left[ (1 -
\gamma_{5}) \psi \right] }{\left[ i\not\partial+e\,\gamma^{\mu} 
(1 -\gamma_{5}) A_{\mu}(\epsilon^{2}-1) \right]}\, + \nonumber \\
& + & \frac{i\,\bar{\psi}^{*}\epsilon^{2}c(i\not\partial)\left[ \bar{\psi}
(1 -\gamma_{5}) \right]}{ \left[ i\not\partial+e\,\gamma^{\mu} 
(1 -\gamma_{5}) A_{\mu}(\epsilon^{2}-1) \right]}.
\een
It can be seen easily that when one take the limit 
$\epsilon^{2} \longrightarrow 1$, the original proper solution of the CSM,
shown below, is obtained.

The final part is the computation of the 
one-loop anomaly of the Chiral Schwinger model.
Firstly, we have to construct some very important matrices,
\beq
S_{A}^{B} = \frac{\delta_{r}\delta_{l}\,S_{BV}}{\delta
\Phi^{B}\,\delta \Phi^{*}_{A}}
\eeq
with the proper solution, the BV action, given by
\ben
S_{BV} &=& \int\,d^{2}x \, \{ \; -  \frac{1}{4} \, F_{\mu\nu} F^{\mu\nu} +
\bar{\psi} i \not\partial \psi + {e \over 2}\bar{\psi}\gamma_{\mu} (1 - \gamma_{5}) 
A^{\mu} \psi  +  A^{*}_{\mu}\partial^{\mu}c  \nonumber \\
&+& i\,\psi^{*}(1 -\gamma_{5})\psi c - i\,\bar{\psi}^{*}\bar{\psi}(1 +\gamma_{5})c \;\}
\een
then
\beq
S_{B}^{A} = \left( \begin{array}{cc}
-ic(1 -\gamma_{5}) & 0 \\
0 & ic(1 +\gamma_{5})
\end{array} \right) .
\eeq

The operator ${\cal I}_{AB}$ in this case is,
\beq
{\cal I}_{AB} = \frac{\delta_{l}\delta_{r} \left[ I(\Phi) +
\Phi^{*}_{c}R^{c}(\Phi) \right]}{\delta \Phi^{A}\delta \Phi^{B}}
\eeq
and the result is,
\beq
{\cal I}_{AB} = \left( \begin{array}{cc}
0 & -\,{e \over 2}\gamma_{\mu} (1 -\gamma_{5}) A^{\mu} \\
{e \over 2}\gamma_{\mu} (1 -\gamma_{5}) A^{\mu} & 0
\end{array} \right) 
\eeq

The one-loop anomaly is given by:
\ben
{\cal A} & \equiv & (\Delta S)_{R} \\
(\Delta S)_{R} & = & \lim_{\Lambda^{2} \rightarrow \infty}
[\Omega_{0}]_{0} \\
\Omega_{0} & = & \left[ \epsilon^{2}S_{A}^{A} \right] +
\left[ \epsilon^{2}S_{B}^{A}{\cal O}^{BC}{\cal I}_{CA} \right] + 
O \left( \frac{(\Phi^{*})^{2}}{\Lambda^{2}} \right)
\een

For the first term
\ben
\epsilon^{2}S_{A}^{A} & = & \epsilon^{2}\,tr\,S_{B}^{A} \nonumber \\
& = & 0
\een
and we have that
\beq
(\Delta S)_{R} = \lim_{\Lambda^{2} \rightarrow \infty} 
tr \left[ \epsilon^{2}S_{B}^{A}{\cal O}^{BC}{\cal I}_{CA} \right]
\eeq

Using the $\gamma$ matrix representation
\beq
\gamma^{0} = \left( \begin{array}{cc}
0 & -1 \\
-1 & 0
\end{array} \right)         
\eeq
\beq
\gamma^{1} = \left( \begin{array}{cc}
0 & -i \\
i & 0
\end{array} \right)         
\eeq
and
\beq
\gamma^{5} = \,-\,i\,\gamma_{1}\,\gamma_{0}
\eeq
in this representation we have that $\gamma_{5}^t=\gamma_{5}$.

Finally, after a little algebra
\beq
(\Delta S)_{R} = \lim_{\Lambda^{2} \rightarrow \infty} 
tr \left[ \epsilon^{2}(-ec)\frac{\epsilon^{2}-1}{\partial^{2}}
(\partial_{\mu}A^{\mu}\,-\,\epsilon^{\mu\nu} \partial_{\mu}A_{\nu} ) \right]
\eeq

But we know that
\ben
& & \lim_{\Lambda^{2} \rightarrow \infty} 
tr \left[ \epsilon^{2}\, F \partial^{n} \, \frac{\epsilon^{2}-1}{\partial^{2}}\,
\partial \, G \, \partial^{m} \right] = \\
& = & \frac{-i}{2 \pi} \left[ \, \sum_{k=0}^{m}
\left( \begin{array}{c}
m \\ k
\end{array} \right)\frac{(-1)^{k}}{n+m+1-k} 
\left( 1-\frac{1}{2^{n+m+1-k}} \right) \right]
\int \, d^{2}x\,F\,\partial^{n+m+1}\,G \nonumber
\een    

In our case
\ben
n & = & m = 0 \nonumber \\
F & = & 2ec \nonumber \\
\partial G & = & \partial_{\mu}A^{\mu}\,-\,\epsilon^{\mu\nu} \partial_{\mu}A_{\nu}   
\een
and the final result is
\beq
{\cal A} = (\Delta S)_{R} = \frac{ie}{2 \pi} 
\int\,d^{2}x\,c\, \left( \partial_{\mu}A^{\mu}\,-\,\epsilon^{\mu\nu} \partial_{\mu}A_{\nu} \right)
\eeq
which is the one-loop anomaly of the CSM.

\section{Conclusions}

The non-local regularization formalism is a new and a quite powerfull
method to regularize theories with a perturbative expansion which
have higher loop order divergences.  The field-antifield framework
exhibits a divergence on the application of the $\Delta$ operator.
Hence it needs a regularization.  The conection
between both generates an extended non-locally regularized  BV quantization 
method.  The quantization of anomalous gauge theories can be
computed exactly.  The one-loop anomaly of the chiral Schwinger model
has been calculated.

\vspace{1cm}

\noindent {\bf Acknowledgment:} the author would like to thank Nelson R. F.
Braga for very valuable discussions. This work is supported by
CAPES (Brazilian Research Agency).  

%\newpage


\vskip 1cm


\vspace{1cm}

\begin{thebibliography}{30}

\bibitem{BV}I. A. Batalin and G. A. Vilkovisky, Phys. Lett. B 102(1981)27, 
Phys. Rev. D 28(1983)2567.
\bibitem{Jon}F. DeJonghe,``The Batalin-Vilkovisky Lagrangian Quantization 
Scheme with Applications to the Study of Anomalies in Gauge Theories",Ph.D. 
thesis K. U. Leuven, hep-th 9403143.
\bibitem{Gomis}J. Gomis, J. Paris and S. Samuel, Phys. Rep. 259(1995)1.
\bibitem{Hen}M. Henneaux, Nucl. Phys.B (Proc. Suppl.) 18 A (1990)47.
\bibitem{ZJ}J. Zinn-Justin, in Trends in Elementary Paritcle Theory,
Lecture notes in Physics 37, Int. Summer Inst. on Theor. Phys., Bonn 1974, 
eds. H. Rollnik and K. Dietz (Springer, Berlin, 1975); Nucl. Phys. 
B 246(1984)246.
\bibitem{Wit}E. Witten, Mod. Phys. Lett. A 5(1990)487; A. Schwarz,
Commun. Math. Phys. 155(1983)249; O. M. Khudaverdian and A. P. Nercessian, 
hep-th 9303136; S. Aoyama and S. Vandoren, hep-th 9305087.
\bibitem{Troost}W. Troost, P. van Nieuwenhuizen and A. van Proyen,
Nucl. Phys. B 333(1990)727.
\bibitem{Pro}A. van Proyen,in Proc. Conf. and Symmetries, 1991, Stony
Brook,  May 20-25, 1991, eds. N. Berkovits et al. (Word Scientific,
Singapure, 1992) p. 388.
\bibitem{Nelson}R. Amorin and N. R. F. Braga, Phys. Rev. D 57 (1998) 1225.
\bibitem{Pauli}W. Pauli and F. Villars, Rev. Mod. Phys. 21(1949)434.
\bibitem{Diaz}A. Diaz, W. Troost, P. van Nieuwenhuizen and A. van Proyen, 
Int. J. Mod. Phys. A 4(198)3959.
\bibitem{Hat}M. Hatsuda, W. Troost, P. van Neuwenhuizen and A. van Proyen, 
Nucl. Phys. B 335(1990)166.
\bibitem{BPHZ}For a pedagogical account see: W. Zimmerman, in Lectures on 
Elementary Particles and Quantum Field Theory, eds. S. Deser, M. Grisary and 
H. Pendleton, MIT Press, Brandeis Lectures, 1970;
J. Lowenstein, ``Seminars on Renormalization Theory", Technical 
Report no. 73-068, 1972, University of Pittsburgh;
M. O. C. Gomes, ``Some Applications of Normal Product 
Quantization in Renormalization Perturbation Theory", Ph.D. Thesis, 
University of Pittsburg, 1972.
\bibitem{Jonghe}F. DeJonghe, J. Paris and W. Troost, Nucl. Phys. B 
476(1996)559.
\bibitem{EU}E. M. C. Abreu and N. R. F. Braga, 
Int. J. Mod. Phys. A (1998) 4249.
\bibitem{Tonin}M. Tonin, Nucl. Phys. B (Proc. Suppl.) 29(1992)137.
\bibitem{NL}D. Evens,J. W. Mofat, G. Kleppe and R. P. Woodard, Phys.
Rev. D 43(1991)499.
\bibitem{Kle}G. Kleppe and R. P. Woodard, Ann. Phys. (NY) 221(1993)106.
\bibitem{Woo}G. Kleppe and R. P. Woodard, Nucl. Phys. B 388(1992)81.
\bibitem{Sch}J. Schwinger, Phys. Rev. 82(1951)664.
\bibitem{Hand}B. J. Hand, Phys. Lett. B 275(1992)419.
\bibitem{Clay}M. A. Clayton, L. Demopoulos and J. W. Moffat, 
Int. J. Mod. Phys. A 9(1994)4549.
\bibitem{Paris}J. Paris, Nucl. Phys. B 450(1995)357.
\bibitem{JR}R. Jackiw and R. Rajaraman, Phys. Rev. Lett. 54 (1985) 1219.
\bibitem{WZ}J. Wess and B. Zumino, Phys. Lett. B 37(1971)95; W. A.
Bardeen and B. Zumino, Nucl. Phys. B 244(1984)421.
\bibitem{Sie}F. DeJonghe, R. Siebelink and W. Troost, Phys. Lett. B
396(1993)295. 
\bibitem{Back}P. Ginsparg, ``Applied Conformal Field Theory",
Lectures at Les Houches Summer School, 1988.


\end{thebibliography}
\end{document}